\documentclass[twocolumn,aps,pre,superscriptaddress,nofootinbib]{revtex4-2}

\usepackage{pifont}
\usepackage[caption=false]{subfig}

\usepackage{graphicx}
\usepackage[dvipsnames,svgnames]{xcolor}
\usepackage{bm,bbm}
\usepackage{braket}
\usepackage{amsthm,amsmath,amssymb}
\usepackage{enumerate}
\usepackage{xcolor}
\usepackage{booktabs,multirow}
\usepackage{mathtools,bbding}
\usepackage[percent]{overpic}
\usepackage{microtype}
\usepackage{array,adjustbox,afterpage}
\usepackage[T1]{fontenc}
\usepackage[utf8]{inputenc}
\usepackage{tikz}
\usetikzlibrary{calc}
\usepackage[english]{babel}
\usepackage{newtxtext,newtxmath}

\usepackage{lipsum}

\usepackage[linesnumbered,ruled]{algorithm2e}
\makeatletter
\renewcommand{\fnum@figure}{FIG. \thefigure} 
\makeatother

\usepackage[bookmarks=false, colorlinks=true, linkcolor=blue, citecolor=purple, urlcolor=purple]{hyperref}
\usepackage{cleveref}
\usepackage[normalem]{ulem}
\usepackage{color,soul}
\usepackage{lipsum}

\Crefname{subfigures}{figure}{figures}
\Crefname{subfigures}{Figure}{Figures}

\def\bra#1{\mathinner{\langle{#1}|}}
\def\ket#1{\mathinner{|{#1}\rangle}}

\newcommand{\xdownarrow}[1]{{\left\downarrow\vbox to #1{}\right.\kern-\nulldelimiterspace}} 

\hypersetup{
    bookmarksnumbered=false, 
    breaklinks=true,
    unicode=false, 
    pdfstartview={FitH}, 
    pdftitle={}, 
    pdfnewwindow=true, 
}

\begin{document}

\title{Geometrical optimization of spin clusters for
the preservation of quantum coherence}
\author{Lea Gassab}
\email{lgassab20@ku.edu.tr}
\affiliation{Department of Physics, Ko{\c{c}} University, 34450 Sar{\i}yer, Istanbul, T\"{u}rkiye}
\author{Onur Pusuluk}
\affiliation{Faculty of Engineering and Natural Sciences, Kadir Has University, 34083, Fatih, Istanbul, T\"{u}rkiye}
\affiliation{Department of Physics, Ko{\c{c}} University, 34450 Sar{\i}yer, Istanbul, T\"{u}rkiye}
\author{\"{O}zg\"{u}r E. M\"{u}stecapl{\i}o\u{g}lu}
\affiliation{Department of Physics, Ko{\c{c}} University, 34450 Sar{\i}yer, Istanbul, T\"{u}rkiye}
\affiliation{T\"{U}B\.{I}TAK Research Institute for Fundamental Sciences, 41470 Gebze, T\"{u}rkiye}
\affiliation{Faculty of Engineering and Natural Sciences, Sabanci University, Tuzla 34956, Istanbul, T\"urk{\.i}ye}

\begin{abstract}

We investigate the influence of geometry on the preservation of quantum coherence in spin clusters subjected to a thermal environment. Assuming weak inter-spin coupling, we explore the various buffer network configurations that can be embedded in a plane. Our findings reveal that the connectivity of the buffer network is crucial in determining the preservation duration of quantum coherence in an individual central spin. Specifically, we observe that the maximal planar graph yields the longest preservation time for a given number of buffer spins. Interestingly, our results demonstrate that the preservation time does not consistently increase with an increasing number of buffer spins. Employing a quantum master equation in our simulations, we further demonstrate that a tetrahedral geometry comprising a four-spin buffer network provides optimal protection against environmental effects.

\end{abstract}

\maketitle

\section{Introduction}
Quantum coherence plays a vital role in a wide range of quantum technologies~\cite{streltsov2017colloquium}, including quantum computing~\cite{unruh1995maintaining,duan1997preserving}, quantum sensing~\cite{RevModPhys.89.035002}, quantum metrology~\cite{Shlyakhov2018QMetrology}, and quantum cryptography~\cite{Gisin2002Crypto}. Moreover, investigating quantum coherence holds tremendous potential for enhancing our understanding of the underlying physics of living systems~\cite{chin2012coherence}. However, the presence of environmental noise and the detrimental effects of decoherence pose substantial challenges \cite{schlosshauer2019quantum,brandt1999qubit}.

Noise can arise from various sources, such as thermal fluctuations, electromagnetic radiation, and interactions with neighboring particles. To address this challenge, several strategies have been proposed. These encompass intentionally introducing supplementary noise during the coupling process \cite{yang2022quantum}, using periodical kicks \cite{viola1999dynamical}, implementing a non-Hermitian driving potential \cite{huang2021effective}, employing correlated channels for interaction \cite{sk2022protecting}, leveraging topological edge states \cite{bahri2015localization,nie2020topologically,yao2021topological}, integrating auxiliary atoms \cite{faizi2019protection}, or even surfaces \cite{bluvstein2019extending} to safeguard coherence \cite{faizi2019protection}.

The existing strategies for protecting coherence in artificial systems predominantly rely on external drives \cite{viola1999dynamical}, the presence of spontaneously occurring coherence \cite{faizi2019protection}, localization on edge states or positioning near surfaces \cite{bluvstein2019extending}, sophisticated interaction control methods \cite{sk2022protecting}, or tailored noise control techniques \cite{yang2022quantum}. In contrast, nature appears to have discovered a simpler solution by leveraging the arrangement and connectivity of molecular networks, as argued in light-harvesting complexes containing chromophores. Although the presence and beneficial effects of quantum coherence in such biological systems are still a subject of debate, we propose a theoretical exploration to identify ideal molecular geometries for coherence protection. By doing so, we aim to guide the engineering of artificial molecules or materials that can store quantum coherence. Our approach holds the potential to shed light on the lifetime and possible existence of quantum coherence in physiological conditions in nature. Our studies may be further complemented by quantum chemistry calculations to assess the energetic stability of these artificial quantum networks and search for their natural analogs.

We have a specific focus on investigating the influence of geometrical degrees of freedom on the preservation of quantum coherence within the core of an atomic cluster, where the atoms are assumed to be two-level systems, modelled as spin-1/2 particles. In order to achieve this objective, we thoroughly examine spin-star networks to evaluate the efficacy of peripheral spins as protective barriers, shielding the central spin from its surrounding environment. The sub-network, comprised of these peripheral spins known as the buffer network, serves as an intermediary layer that adeptly absorbs and dissipates environmental noise, while simultaneously upholding the coherence of the central spin. The utilization of buffer networks holds significant potential for applications in quantum computation, quantum sensing, and the study of biological molecules \cite{thatcher1993phosphonates,fisher2015quantum}. Our findings carry practical implications for various applications that demand long-lived coherent spin states, including quantum memories \cite{gentile2017optically}, quantum magnetometry \cite{barry2020sensitivity}, quantum control and computation \cite{vandersypen2001experimental,vandersypen2005nmr}, as well as biomedical imaging \cite{gossuin2010physics}.

\begin{figure}[t!]
	\centering
        \includegraphics[width=0.32\textwidth]{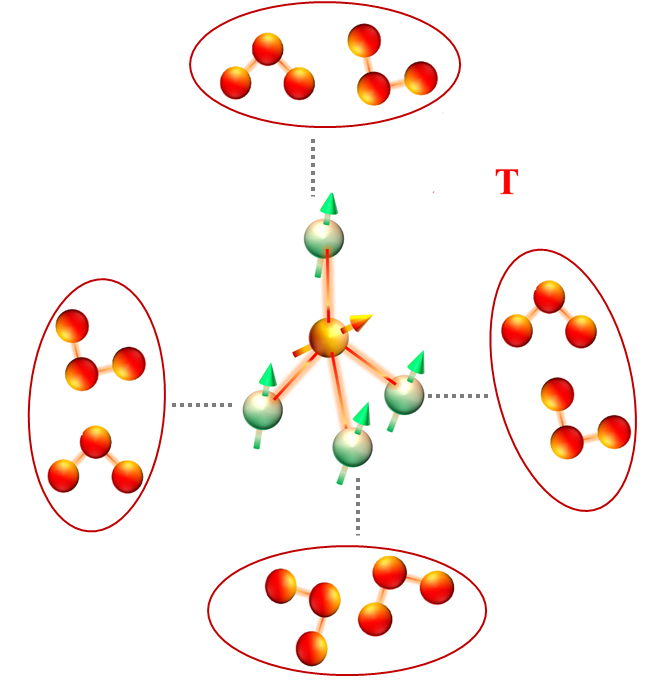}
        \caption{A central spin surrounded by four buffer spins, each one in its own thermal environment at temperature $T$. To illustrate how the central spin has negligible interaction with the environment, while the buffer spins can have frequent interactions, the thermal environment is depicted as a molecular bath. This system can be effectively viewed as a composite open quantum system, where the central spin is isolated from the environment, and the buffer spins each has their own local thermal dissipation channel.}

       \label{Fig::SpinNetwork}
\end{figure}

The paper is structured as follows: Sec.~\ref{sec:model} presents an overview of the model employed in this study. In Sec.~\ref{sec:results}, we examine buffer networks with varying numbers of spins and connectivities, and present our findings. Lastly, in Sec.~\ref{sec:conclusion}, we summarize our conclusions based on the results obtained.

\section{Model System}\label{sec:model}

\subsection{Spin-Star Network with Different Topologies}

\begin{figure}
    \begin{center}
        \centering
        \includegraphics[width=0.45\textwidth]{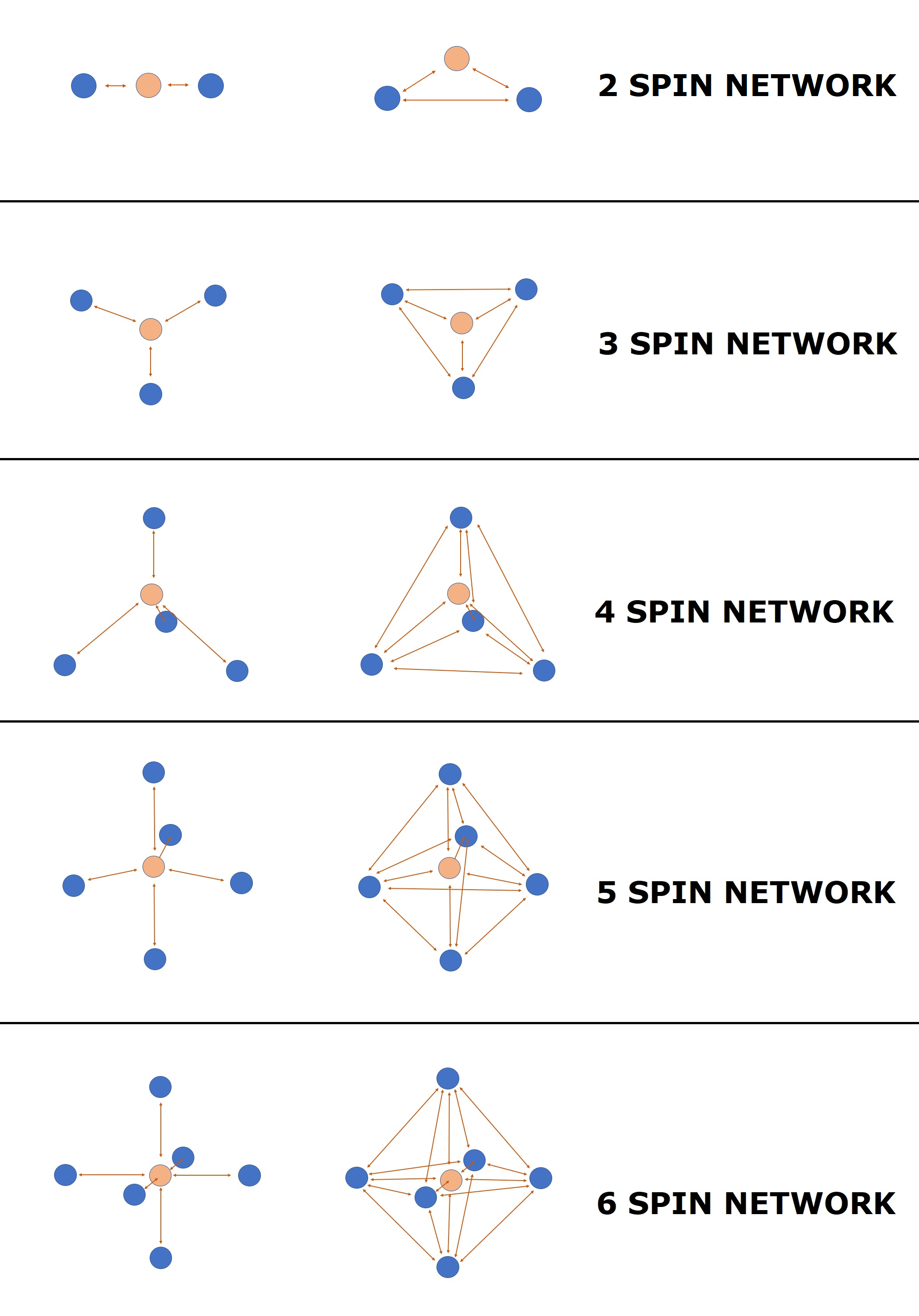}
        \caption{Depiction of extreme geometries for central spin-buffer spins network for $N=2,3,4,5,6$ buffer spins. The central spin, depicted in orange (light) color, is isolated from the environment, yet it is coupled to the blue (dark) colored buffer spins, which have local thermal dissipation pathways. For a particular number of buffer spins, $N$, we consider all the feasible buffer networks that can be embedded in a plane. Two extreme cases are depicted here: (Left column) No connectivity and (Right column) maximum connectivity within the  buffer spin network.}
        \label{Fig::Geometries}
    \end{center}
\end{figure}

Consider a cluster of $N+1$ spins that consists of a central spin surrounded by a buffer network, as illustrated in Fig.~\ref{Fig::SpinNetwork}. Each buffer spin is individually connected to a thermal bath, while the central spin does not engage in direct interactions with the environment. Our objective is to analyze the impact of buffer network size and topology on the coherence of the central spin. To achieve this, we focus on buffer networks that can be represented as planar graphs. We explore various scenarios by considering different numbers of buffer spins, ranging from the two extremes: (i) no coupling between buffer spins, and (ii) pairwise interactions between all nearest neighbor buffer spins (see Fig.~\ref{Fig::Geometries}).

For the sake of simplicity, we assume that all spins are identical with a magnitude of $1/2$. When subjected to a magnetic field aligned along the quantization axis $z$, an energy difference denoted by $\hbar \omega$  emerges between the lower state $\ket{0}$  (spin-up) and the upper state $\ket{1}$  (spin-down), enabling a two-level description.

The buffer spins are brought into thermal equilibrium with the surrounding environment at an inverse temperature \(\beta\), while the central spin is initially prepared in a maximally coherent superposition state. Then, the initial cluster state can be represented as a product state,
\begin{equation}
\rho(0) = \frac{\ket{0}+\ket{1}}{\sqrt{2}}\otimes \rho_{th}^{\otimes n}.
\end{equation}
We take $\hbar$=1. The thermal state of the buffer spins is defined as
\begin{equation}
\rho_{th} = \frac{\mathrm{e}^{- \beta \frac{\omega}{2}\hat\sigma_{z}}}{\mathcal{Z}},
\end{equation}
where \(\sigma_{z} = \ket{0} \bra{0} - \ket{1} \bra{1}\) represents the Pauli-\(z\) operator and \(\mathcal{Z} = 2 \cosh[\beta \, \omega/2]\,\) denotes the partition function. Here, we assume the Boltzmann constant \(k_B\) to be equal to 1. The total Hamiltonian of the spin cluster reads as an XX central spin model \cite{ccakmak2017thermal,manzano2012quantum,ryan2021quantum,turkpencc2017photonic,militello2011genuine}, which is a simplified case of general Richardson-Gaudin spin cluster models \cite{gaudin1976diagonalization},
\begin{equation}\label{Eq::Hamiltonian}
   \hat  H =  \sum_{i=1}^{N+1} \frac{\omega}{2}\hat\sigma_{z}^{(i)} + \sum_{i\neq j} g_{ij} (\hat\sigma_{x}^{(i)}\hat\sigma_{x}^{(j)}+\hat\sigma_{y}^{(i)}\hat\sigma_{y}^{(j)}) \, .
\end{equation}
We take $\hbar = 1$. The Pauli spin-$1/2$ operators for the $i^{th}$ spin are denoted by $\sigma_{x}^{(i)}$, $\sigma_{y}^{(i)}$, and $\sigma_{z}^{(i)}$, and the interaction strength between the spin pair $(i,j)$ is represented by $g_{ij}$. The central spin, labeled by ``1'', is coupled to the buffer spins at strength $g_{1j} = g\neq 0$ in all geometries under consideration. On the other hand, each buffer network is defined by a different array of coupling constants consisting of zero or \(g\) values, which can be represented by a planar graph.

\begin{table*}[t!]
    \centering
        \caption{The dimensionless time for which the relative entropy \(S(\cdot\|\cdot)\), trace distance \(T(\cdot\|\cdot)\), relative entropy of coherence \(C_{RE}(\cdot)\) or \(L_1\) norm of coherence \(C_{L_1}(\cdot)\) become always inferior to $10^{-4}$ at the center of spin cluster for the parameters $\omega = 1$, $g = 0.002$, $\gamma = 0.0005$, and $T = 0.4$. The left and right columns compare the same quantity for the left and right geometries depicted in Fig.~\ref{Fig::Geometries}, which correspond to vanishing and maximal connectivities in the buffer network, respectively. The initial state of the buffer spins is a maximally coherent superposition state.}
    \begin{tabular}{|p{1cm}|p{1.5cm}|p{1.5cm}|p{1.5cm}|p{1.5cm}|p{1.5cm}|p{1.5cm}|p{1.5cm}|p{1.5cm}|}
    \hline
        \(N+1\) &\multicolumn{2}{|c|}{$S(\rho_1 \| \rho_{th})$} & \multicolumn{2}{|c|}{$T(\rho_1, \rho_{th})$} & \multicolumn{2}{|c|}{$C_{RE}(\rho_1)$} & \multicolumn{2}{|c|}{$C_{L_1}(\rho_1)$}  \\ \hline
        3 & 20640   & 24210  & 20630 & 23710   & 38970 & 46680 & 41770 & 50150\\ \hline
        4 & 17270 & 29870 & 16870 & 29730 & 32310 & 58320 & 34590 & 62020 \\ \hline
        5 & 14990 & \textbf{32930} & 14250 & \textbf{32360} & 27250 & \textbf{64660} & 29600 & \textbf{66870} \\ \hline
        6 & 13460 & 25440 & 12410 & 24800 & 24410 & 48070 & 25820 & 52010 \\ \hline
        7 & 12930  & 19410 & 11030 & 17780 & 23260 & 35320 & 23260 & 37450 \\ \hline
    \end{tabular}
    \label{Table::Times}
\end{table*}

\begin{table*}[t!]
    \centering
        \caption{The mean value of the relative entropy \(S(\cdot\|\cdot)\), trace distance \(T(\cdot\|\cdot)\), relative entropy of coherence \(C_{RE}(\cdot)\) or \(L_1\) norm of coherence \(C_{L_1}(\cdot)\) of the central spin from time $t_1 = 29000 $ to time $t_2 = 30000 $ for the parameters $\omega = 1$, $g = 0.002$, $\gamma = 0.0005$, and $T = 0.4$. The left and right columns compare the same quantity for the left and right geometries depicted in Fig.~\ref{Fig::Geometries}, which correspond to vanishing and maximal connectivities in the buffer network, respectively.}
    \begin{tabular}{|p{1cm}|p{1.5cm}|p{1.5cm}|p{1.5cm}|p{1.5cm}|p{1.5cm}|p{1.5cm}|p{1.5cm}|p{1.5cm}|}
    \hline
        \(N+1\) &\multicolumn{2}{|c|}{$S(\rho_1 \| \rho_{th})$} &\multicolumn{2}{|c|}{$T(\rho_1, \rho_{th})$}  & \multicolumn{2}{|c|}{$C_{RE}(\rho_1)$} & \multicolumn{2}{|c|}{$C_{L_1}(\rho_1)$}  \\ \hline
        3 & 1.13$\times10^{-6}$ & 1.46$\times10^{-5}$ & 1.12$\times10^{-6}$ & 1.33$\times10^{-5}$ & 4.60$\times10^{-4}$ & 1.80$\times10^{-3}$ & 9.17$\times10^{-4}$ & 3.54$\times10^{-3}$ \\ \hline
        4 & 9.60$\times10^{-8}$ & 1.12$\times10^{-4}$ & 9.35$\times10^{-8}$ & 1.07$\times10^{-4}$ & 1.36$\times10^{-4}$ & 5.07$\times10^{-3}$ & 2.68$\times10^{-4}$ & 1.01$\times10^{-2}$ \\ \hline
        5 & 8.26$\times10^{-9}$ & \(\mathbf{2.54 \times 10^{-4}}\) & 6.73$\times10^{-9}$ & \(\mathbf{2.14 \times 10^{-4}}\) & 3.89$\times10^{-5}$ & \(\mathbf{7.38 \times 10^{-3}}\) & 7.25$\times10^{-5}$ & \(\mathbf{1.42 \times 10^{-2}}\) \\ \hline
        6 & 1.83$\times10^{-9}$ & 2.47$\times10^{-5}$ & 5.04$\times10^{-10}$ & 2.07$\times10^{-5}$ & 1.54$\times10^{-5}$ & 2.29$\times10^{-3}$ & 1.94$\times10^{-5}$ & 4.41$\times10^{-3}$ \\ \hline
        7 & 1.35$\times10^{-9}$ & 9.41$\times10^{-7}$ & 5.32$\times10^{-11}$ & 4.35$\times10^{-7}$ & 1.17$\times10^{-5}$ & 3.89$\times10^{-4}$ & 6.32$\times10^{-6}$ & 6.39$\times10^{-4}$ \\ \hline
    \end{tabular}
    \label{Table::Value2}
\end{table*}


That is to say, the coupling constants between buffer spins can be drawn in the plane so that buffer spins and their non-zero coupling constants are represented by vertices and edges, respectively, and no two edges of the resulting graph intersect at a point other than a vertex. The maximum number of edges between the vertices corresponding to the buffer spins in such a graph is $3N-6$, in the case of \(N \geq 3\) buffer spins. This means that the array of coupling constants between buffer spins cannot include more than \(3\,N - 6\) non-zero values. Hence, the number of possible geometries for a cluster of \(N+1\) spins becomes
\begin{equation}
M = \sum_{k=0}^{3\,N - 6} \begin{pmatrix} E \\ k \end{pmatrix} \, ,
\end{equation}
where $E=N(N-1)/2$ is the number of edges for a complete graph.
Fig.~\ref{Fig::Geometries} shows two extreme geometries, contributing to this sum, for  $N=2, 3, 4, 5$ and $N=6$ buffer spins representing the buffer network; $k$ classifies the buffer network geometries with respect to the number of edges in them. The left geometry corresponds to the \(k=0\) term, whereas the right geometry corresponds to one of the \(k = 3\,N - 6\) terms, which allows putting the central spin right in the middle of the bulk network.

\subsection{Open System Dynamics}
\begin{figure*}[t!]
    \begin{center}
        \centering
        \includegraphics[width=0.7\textwidth]{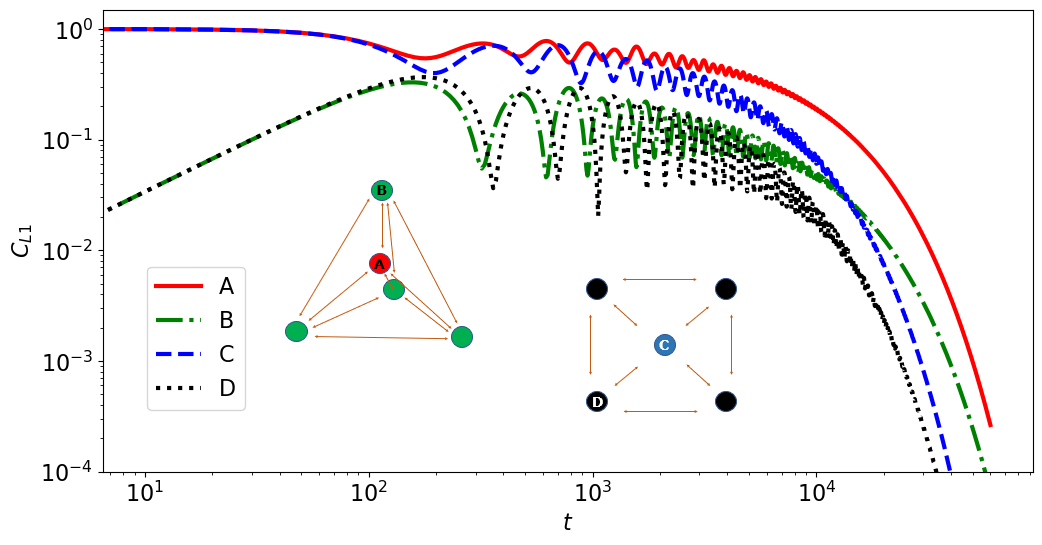}
        \caption{The behaviour of \(L_1\) norm of coherence \(C_{L_1}(\cdot)\) for four-buffer spin networks with respect to dimensionless time $t$ for the parameters $\omega = 1$, $g = 0.002$, $\gamma = 0.0005$, and $T = 0.4$. We use Logarithmic scales for both vertical and horizontal axis. Two geometries are compared here. The first one is the tetrahedral buffer network which seems the best candidate to preserve coherence according to the results in Tables~\ref{Table::Times}~and~\ref{Table::Value2}. The second one is a square buffer network, which is another planar graph with one less edge than the first. The solid (red) and the dash-dotted (green) curves illustrate the quantum coherence found in the central (A) and buffer (B) spins in the tetrahedral geometry. The dashed (dark blue) and pointed (black) curves represent the quantum coherence localized in the central (C) and buffer (D) spins in the square geometry.}
        \label{Fig::CohIn4Spin}
    \end{center}
\end{figure*}


We describe the open quantum system dynamics of the spin-star network by the following Lindblad master equation~\cite{breuer2002theory},
\begin{equation} \label{Eq::MasterEq}
\dot{\rho}(t) = -i[\hat H,\rho(t)] + \mathcal{D}(\rho(t)) \, ,
\end{equation}
where \(\hbar\) is taken to be 1 and the unitary contribution to the dynamics is provided by the self-Hamiltonian of the system given in Eq.~\eqref{Eq::Hamiltonian}.

By assuming weakly coupled buffer spins, local thermal dissipation channels are described by the dissipator in Eq.~(\ref{Eq::MasterEq}), which reads
\begin{eqnarray} \begin{aligned} \label{Eq::Dissip}
\mathcal{D}(\rho) &=  \sum_{i=2}^{N+1} \gamma_{i} \, (1+n(\omega))[\hat\sigma_i^-\rho(t)\hat\sigma_i^{+} -\frac{1}{2} \left\{\hat\sigma_i^{+}\hat\sigma_i^-,\rho(t)\right\}]\\
& +\sum_{i=2}^{N+1} \gamma_{i} \, n(\omega) \, [\hat\sigma_i^{+}\rho(t)\hat\sigma_i^- -\frac{1}{2} \left\{\hat\sigma_i^- \hat\sigma_i^{+},\rho(t)\right\}] \, ,
\end{aligned}
\end{eqnarray}
where $n(\omega)$ is the Planck distribution at the spin resonance frequency $\omega$, $\hat\sigma^\pm$ are the Pauli spin ladder operators, and $\gamma_i = \gamma$ is the coupling constant between the environment and $i^{th}$ buffer spin, taken to be homogeneous for each buffer spin independent of the network structure for simplicity.

To determine whether the dissipator~(\ref{Eq::Dissip}) achieves local thermalization in our simulations, we compare the reduced spin states \(\rho_i = \mathrm{tr}_{j_1\cdot\cdot\cdot j_n}[\rho]\), \(j_k \neq i\) with the local thermal state \(\rho_{th}\) using two different measures. The first measure is the relative entropy, defined as:
\begin{equation}\label{Eq::RE}
    S(\rho \| \sigma) = \mathrm{tr}[\rho \, \log_2 \rho] \, - \mathrm{tr}[ \rho \, \log_2 \sigma] .
\end{equation}

The second measure is the trace distance, which can be reduced to
\begin{equation}\label{Eq::TD}
    T(\rho , \sigma) = \frac{1}{2} \mathrm{tr}\left[\sqrt{(\rho - \sigma)^2}\right] \, ,
\end{equation}
for the Hermitian matrices \(\rho\) and \(\sigma\).

For a given number of buffer spins $N$, we investigate the open system dynamics of $M$ different spin clusters and search for the geometry that optimizes the protection of the central spin coherence. For this aim, we quantify the quantum coherence possessed in the reduced state of the central spin \(\rho_1 = \mathrm{tr}_{2\cdot\cdot\cdot n+1}[\rho]\) by the $l_1$ norm of coherence~\cite{baumgratz2014quantifying} that equals to the sum of the magnitude of all the off-diagonal elements of a given density matrix,
\begin{equation}\label{Eq::l1Norm}
    C_{L_1}(\rho)=\sum_{i\ne j}|\rho_{ij}| \, .
\end{equation}

$C_{L_1}$ neglects the signs of distinct coherences in the basis of \(\{\ket{0},\ket{1}\}\) and takes them into account independently of each other. Another measure of coherence that we utilize to determine the amount of coherence in the central spin is the relative entropy of coherence~\cite{baumgratz2014quantifying},
\begin{equation}\label{Eq_REofCoh}
C_{RE}\!\left[\rho\right] = \min_{\varsigma \in
\mathrm{IC}} \left(S\left[\rho \| \varsigma\right]\right) =
S\left[\rho \|  \rho_{d}\right] ,
\end{equation}
where the minimum is taken over the set of incoherent states (IC) that are diagonal in the basis \(\{\ket{0},\ket{1}\}\) and $\rho_{d}$ is the diagonal part of the density matrix $\rho$. $C_{RE}$ measures the distinguishability of a density matrix with a modified copy which is subjected to a full dephasing process.

\section{Results}\label{sec:results}
\subsection{Simulation Results}

The simulations were performed by using scientific Python packages along with key libraries from QuTiP~\cite{johansson2012qutip}. The spin transition frequency $\omega$ was taken as the time and energy scale (such that $\omega = 1$) and dimensionless scaled parameters are used in the simulations. Particularly, inter-spin coupling strength \(g\) and bath dissipation rate \(\gamma\) were taken to be 0.002 and 0.0005 respectively. The temperature of the environment ($T = \beta^{-1}$) was taken as 0.4. The simulations were continued until the quantum coherence of the central spin vanished. Specifically, we record the time after which the quantum coherence is always inferior to $10^{-4}$. The buffer spins are in a thermal initial state at the same temperature as their local bath ($T=0.4$).

Our simulations showed that the dissipator~(\ref{Eq::Dissip}) provides local thermalization for both central and buffer spins. For this aim, we compared the steady-state spin states \(\rho_i(\infty)\) with the local thermal state \(\rho_{th}\) using the relative entropy and trace distance. Additionally, we identified the thermalization time at which the relative entropy between the reduced spin states and \(\rho_{th}\) diminishes. The findings regarding the central spin are summarized in the first two columns of Table~\ref{Table::Times} for the geometries depicted in Fig.~\ref{Fig::Geometries}. The table also presents the times at which the relative entropy of coherence vanishes. Notably, the disappearance of $C_{RE}(\rho_1)$ consistently occurs later than $S(\rho_1 \| \rho_{th})$, indicating that complete decoherence occurs after the populations have reached equilibrium values.

\begin{figure*}[t!]
    \begin{center}
        \centering
        \includegraphics[width=0.7\textwidth]{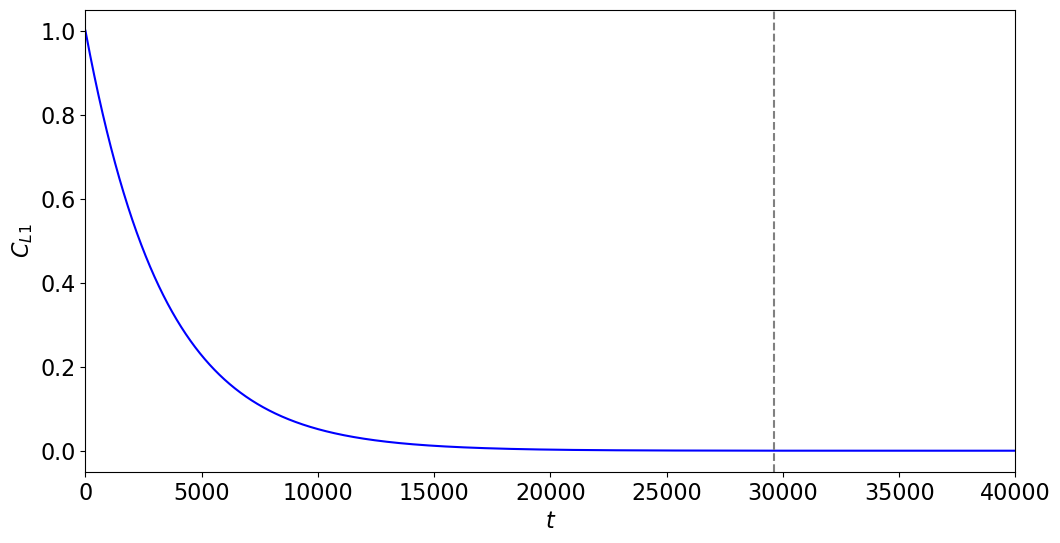}
        \caption{The behavior of \(L_1\) norm of coherence \(C_{L_1}(\cdot)\) for a single spin connected to a thermal bath with respect to dimensionless time $t$. The initial state of the single spin is a maximally coherent superposition state. The time for which the \(L_1\) norm of coherence becomes inferior to $10^{-4}$ is given by the vertically dotted gray line. The parameters are taken as $\omega = 1$ and $T = 0.4$. }
        \label{Fig::Coh1Spin}
    \end{center}
\end{figure*}

Our findings demonstrate a correlation between the protection time of quantum coherence in the central spin and the connectivity of the buffer network, considering a specific number of buffer spins. We observe that the network with maximal planar graph embedding yields the longest protection time. To illustrate this, Table~\ref{Table::Times} presents a comparison of the protection time for different cluster sizes, depicted in Fig.~\ref{Fig::Geometries}, with left and right geometries representing minimal and maximal connectivity in the buffer network, respectively. Additionally, Fig.~\ref{Fig::CohIn4Spin} depicts the time-dependent behavior of central spin quantum coherence in four-buffer spin networks with tetrahedral and square geometries. Notably, the former exhibits a longer survival time for quantum coherence. Thus, our conclusion is that a higher number of buffer spin interactions results in a more effective protective shell around the central spin.

In the scenario of vanishing connectivity, we observe a continuous decrease in quantum coherence as the number of buffer spins increases. This is expected since each buffer spin introduces an additional local thermalization channel, thereby accelerating the thermalization process. However, the behavior of protection time differs in the maximal connectivity scenario. It does not increase monotonically with the number of buffer spins in the cluster. The optimal protection against environmental decoherence, as indicated in the last column of Table~\ref{Table::Times}, is achieved by a four-spin buffer network with a tetrahedral geometry.

Furthermore, Table~\ref{Table::Value2} provides the mean value of quantum coherence in the central spin. The mean value is calculated between $t_1$=29000 and $t_2$=30000, which corresponds to the thermalization time of a single spin initially in a maximally coherent superposition state within a thermal bath, as depicted in Fig.\ref{Fig::Coh1Spin}. According to Table~\ref{Table::Value2}, the tetrahedral geometry with four buffer spins yields the highest mean coherence value. The oscillations appearing in Fig.\ref{Fig::CohIn4Spin} arise from the closed system dynamics. The inter-spin interaction affects the coherence of the central spin periodically. When we connect buffer spins to their local heat baths, these oscillations diminish over time, reducing their amplitudes.
We have also checked the case of the pure dephasing channel and the conclusion remains the same (see Appendix~\ref{apB}).
Finally, alternative combinations of the coupling constant and dissipative rate were explored, as depicted in Appendix~\ref{apC}.

\subsection{Physical Mechanism of Coherence Protection}\label{apA}

\begin{figure*}[t!]
		\centering
		\subfloat[]{\includegraphics[width=0.45\linewidth]{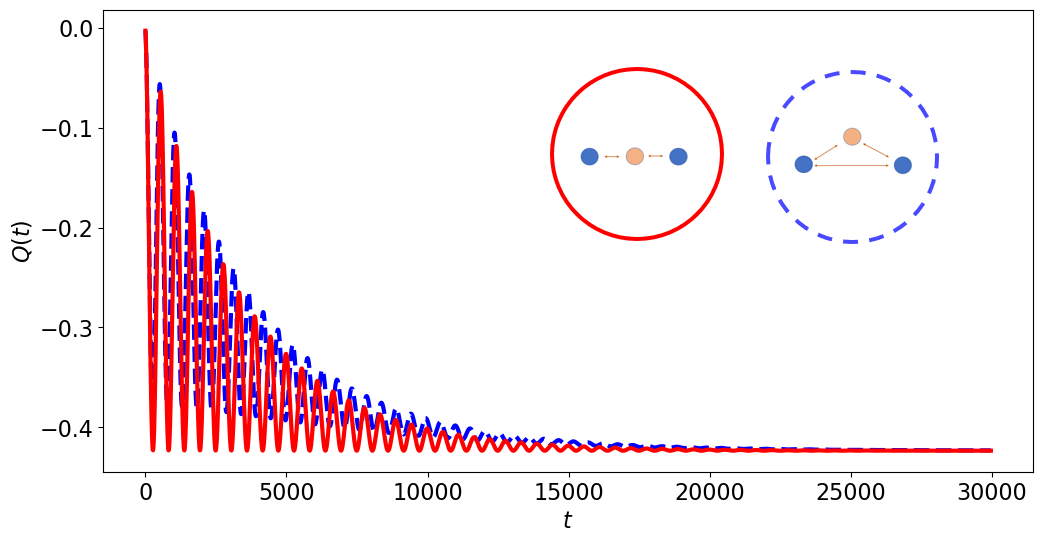}  \label{fig5:a}}
		\quad
		\subfloat[]{\includegraphics[width=0.45\linewidth]{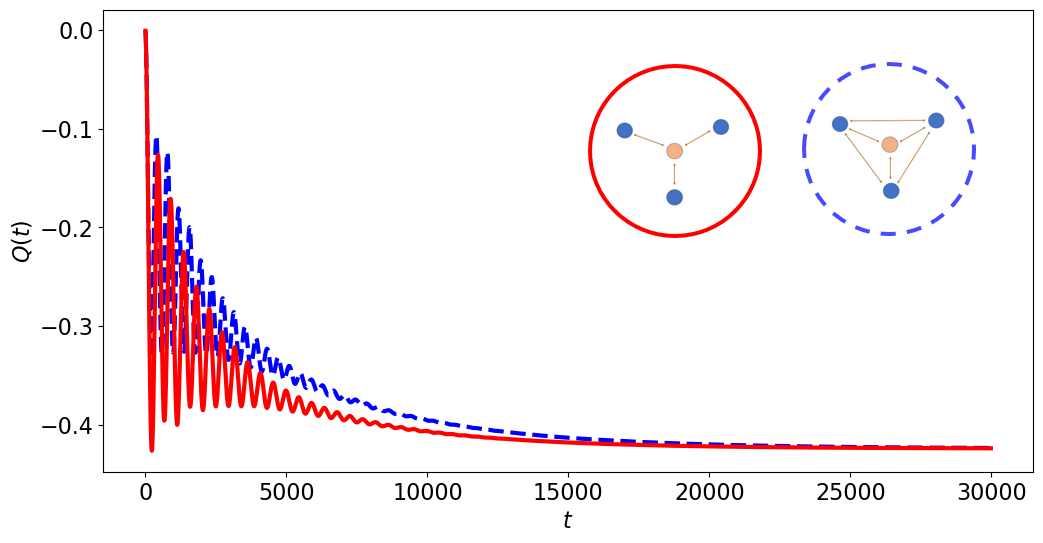}  \label{fig5:b}}
		\hfil
		\subfloat[]{\includegraphics[width=0.45\linewidth]{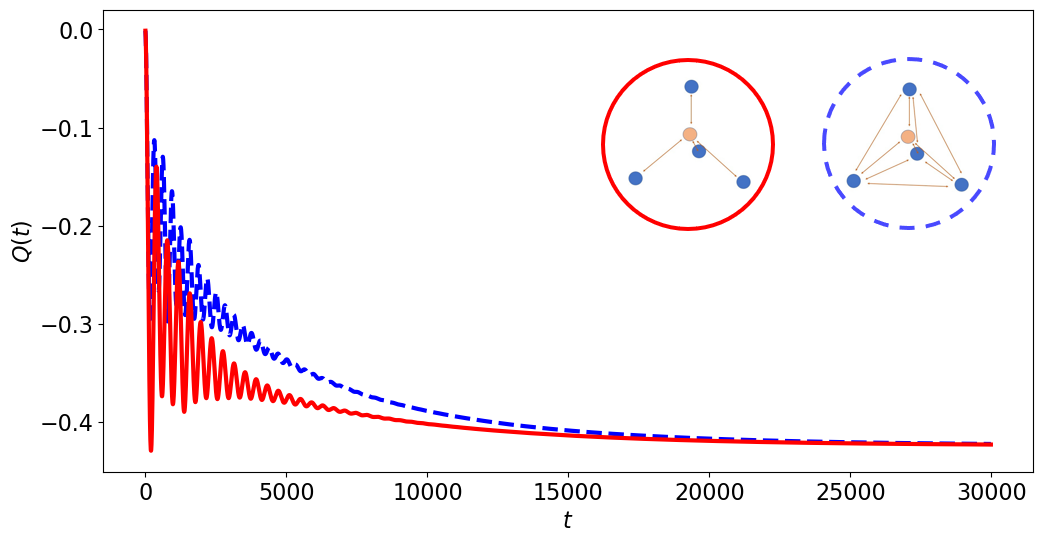}  \label{fig5:c}}
  		\quad
		\subfloat[]{\includegraphics[width=0.45\linewidth]{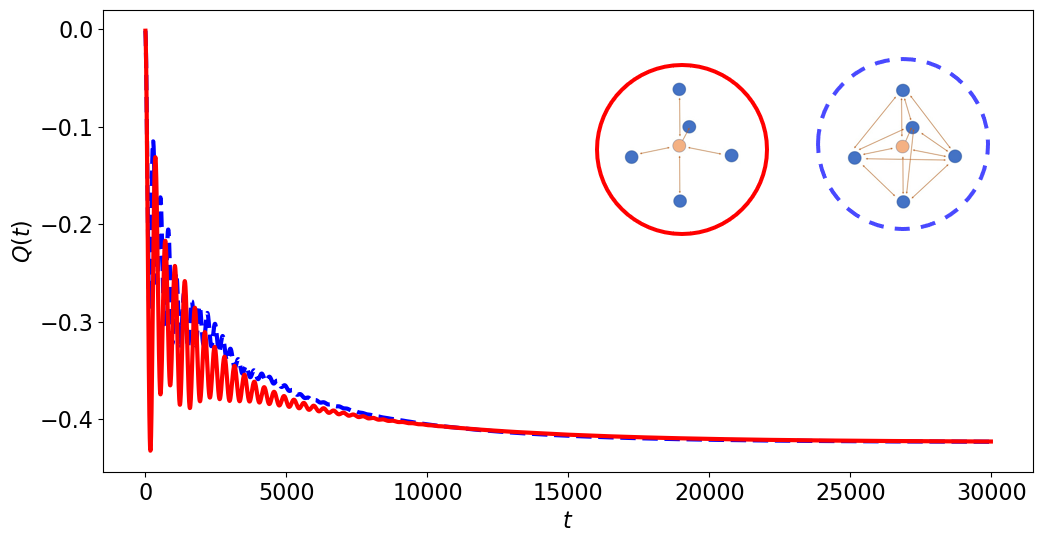}  \label{fig5:b}}
		\caption{Heat current as defined in Eq.~\ref{heat} with respect to dimensionless time $t$. The solid red curve corresponds to the case in which there is no coupling between the buffer spins. The dotted blue curve illustrates the case with coupling between buffer spins. (a) Two buffer spin network. (b) Three buffer spin network. (c) Four buffer spin network. (d) Five buffer spin network.}
		\label{Fig5}
\end{figure*}

The erasure of quantum coherence has been a subject of extensive discussion in the context of quantum computation, primarily as a means to efficiently reset quantum bits before executing quantum algorithms \cite{landauer1961dissipation,bennett1982thermodynamics,piechocinska2000information,shizume1995heat,miller2020quantum}. Landauer's pioneering ideas and erasure protocol have undergone both experimental and theoretical scrutiny \cite{croucher2018information,buffoni2023cooperative,scandi2022minimally}. In this study, we focus on the simplest scenario in which the coherence of a qubit diminishes due to the process of thermalization, similar to the scenario outlined in reference \cite{wang2023fundamental}.

Information erasure, often referred to as `bit reset', represents a critical operation occurring between successive thermodynamic cycles, enabling feedback control of subsystems. Nevertheless, it is imperative to acknowledge that this operation entails a significant energy exchange between the subsystem and its surrounding thermal bath, a concept consistent with Landauer's principle. This principle sets a lower bound on the amount of heat dissipation required for information erasure, thus establishing a close connection between the energy cost and the reduction in entropy within the subsystem.

Wang's work \cite{wang2023fundamental} introduces an inequality for the dissipated energy, which imposes a more stringent lower limit than Landauer's classical formulation. This further underscores the energy-intensive nature of information erasure, frequently surpassing the information gain divided by the inverse of the temperature. It is noteworthy that a more rapid decline in coherence results in a heightened heat transfer during the erasure process.

In our specific system, we can make a comparative analysis of energy transfer between a configuration lacking coupling between buffer spins and one that incorporates such coupling. Our investigation centers on the assessment of the local energy current within our system, which is defined as follows,
\begin{equation}
\label{current}
J(t) = \mathrm{tr}\left(H_s \frac{d}{dt}\rho_s(t)\right).
\end{equation}

In this equation, $H_s= \omega \sigma_z/2$ represents the Hamiltonian and $\rho_s$ the density matrix for the central spin. The total heat exchange in and out of the central spin can be calculated by
\begin{equation}
\label{heat}
Q(t)= \int_{0}^{t}J(\tau) \mathrm{d}\tau.
\end{equation}

To illustrate how the buffer-central spins network structure influences the coherence protection time in the central spin, we consider the different geometries of Fig.~\ref{Fig::Geometries}. Fig.~\ref{Fig5} exhibits a time delay in the exchange of the same amount of heat with the central spin when the buffer spins are coupled. The time delay in exchanging same amount of heat energy in the transient regime results in a delay in achieving the minimum energy transfer essential for erasing the coherence within the central qubit. In our specific case, the energy cost ($E_\mathrm{c}$) of coherence erasure is given by
\begin{equation}
E_\mathrm{c}=\mathrm{tr}\left(H_s (\rho_{\mathrm{fin}}-\rho_{\mathrm{int}})\right)=-0.42,
\end{equation}
where $\rho_{\mathrm{int}}$ and $\rho_{\mathrm{fin}}$ are respectively the initial and final density matrix of the central spin \cite{wang2023fundamental}. As it can be observed in Fig.~\ref{Fig5} all the heat exchange curves corresponding to the different geometries are converging to $E_\mathrm{c}$.
The current analysis serves to elucidate the underlying physical principles supporting our approach to preserving coherence. From an analytical standpoint, it remains an open question to determine which network geometry, involving buffer and central spins, is most effective in delaying the heat exchange required to erase the coherence within the central spin. However, we have systematically conducted a numerical investigation, and the tetrahedral geometry has emerged as the most favorable choice.

\section{Conclusion}
\label{sec:conclusion}

In this study, we have examined the influence of buffer network size and topology on the preservation of quantum coherence in a central spin. Our analysis has specifically concentrated on weak inter-spin coupling, considering buffer networks that can be embedded in a plane without intersecting edges. The results have yielded a noteworthy observation: the preservation time of quantum coherence does not exhibit a consistent increase with the addition of more buffer spins in the cluster. Remarkably, we have identified that a four-spin buffer network, characterized by maximum connectivity and adopting a tetrahedral geometry, provides the most effective means of preserving quantum coherence against the perturbations arising from the thermal environment. This finding highlights the significance of carefully optimizing the buffer network's structural features to enhance the protection of quantum coherence in practical applications.

Notably, this tetrahedral geometry is frequently observed in natural molecules, such as water-ice systems in hexagonal phases~\cite{raza2011proton}, magnetic spin-ice substances~\cite{bramwell2001spin}, and phosphate molecules in Posner's clusters~\cite{fisher2015quantum}. It is worth mentioning that our simplified model does not fully account for such complex molecules. Nevertheless, the correspondence between the tetrahedral network and the optimal geometry holds potential significance for both understanding biochemical processes and advancing artificial quantum technologies.

\newpage
\nocite{*}
%

\onecolumngrid
\newpage
\appendix

\section{Alternative Noise Model }\label{apB}

The case of the pure dephasing channel has also been checked. The dissipator in Eq.~\ref{Eq::Dissip} has been changed to
\begin{eqnarray} \begin{aligned} \label{Eq::Dephase}
\mathcal{D}(\rho) &=
 \sum_{i=2}^{N+1} \gamma_{\mathrm{d}} \,[\hat\sigma_i^z\rho(t)\hat\sigma_i^z -\rho(t)],
\end{aligned}
\end{eqnarray}
where $\gamma_\mathrm{d}$ is the dephasing rate. We considered different dephasing rates. The Tables~\ref{Table::Timesdis} and \ref{Table::Value2dis} are reported for $\gamma_\mathrm{d}=0.00059$. The conclusion of tetrahedral geometry of buffer qubits offering the best protection against decoherence remained robust in case of a dephasing environment as well.

\begin{table*}
    \centering
        \caption{The dimensionless time for which the relative entropy \(S(\cdot\|\cdot)\), trace distance \(T(\cdot\|\cdot)\), relative entropy of coherence \(C_{RE}(\cdot)\) or \(L_1\) norm of coherence \(C_{L_1}(\cdot)\) become always inferior to $10^{-4}$ at the center of spin cluster for the parameters $\omega = 1$, $g = 0.002$, $\gamma_{\mathrm{d}} = 0.00059$, and $T = 0.4$. The left and right columns compare the same quantity for the left and right geometries depicted in Fig.~\ref{Fig::Geometries}, which correspond to vanishing and maximal connectivity in the buffer network, respectively. The initial state of the buffer spins is a maximally coherent superposition state. The results are for the pure dephasing channel.}
    \begin{tabular}{|p{1cm}|p{1.5cm}|p{1.5cm}|p{1.5cm}|p{1.5cm}|p{1.5cm}|p{1.5cm}|p{1.5cm}|p{1.5cm}|}
    \hline
        \(N+1\) &\multicolumn{2}{|c|}{$S(\rho_1 \| \rho_{th})$} & \multicolumn{2}{|c|}{$T(\rho_1, \rho_{th})$} & \multicolumn{2}{|c|}{$C_{RE}(\rho_1)$} & \multicolumn{2}{|c|}{$C_{L_1}(\rho_1)$}  \\ \hline
        3 & 6890  & 9990  & 6890 & 9990  & 13600 & 20440 & 15120 & 22210 \\ \hline
        4 & 6930  & 13600 & 6930 & 13600 & 13780 & 27160 & 15090 & 29520 \\ \hline
        5 & 7130 & \textbf{16960} & 7130 & \textbf{16960} & 13830 & \textbf{33770} & 15010 & \textbf{36700} \\ \hline
        6 & 7060 & 13540 & 7060 & 13540 & 13760 & 27380 & 14840 & 29810 \\ \hline
        7 & 6800  & 13020 & 6800 & 13020 &13540 & 26530 & 14800 & 28900 \\ \hline
    \end{tabular}
    \label{Table::Timesdis}
\end{table*}

\begin{table*}
    \centering
        \caption{The mean value of the relative entropy \(S(\cdot\|\cdot)\), trace distance \(T(\cdot\|\cdot)\), relative entropy of coherence \(C_{RE}(\cdot)\) or \(L_1\) norm of coherence \(C_{L_1}(\cdot)\) of the central spin from time $t_1 = 29000 $ to time $t_2 = 30000 $ for the parameters $\omega = 1$, $g = 0.002$, $\gamma_{\mathrm{d}} = 0.00059$, and $T = 0.4$. The left and right columns compare the same quantity for the left and right geometries depicted in Fig.~\ref{Fig::Geometries}, which correspond to vanishing and maximal connectivities in the buffer network, respectively. The results are for the pure dephasing channel.}
    \begin{tabular}{|p{1cm}|p{1.5cm}|p{1.5cm}|p{1.5cm}|p{1.5cm}|p{1.5cm}|p{1.5cm}|p{1.5cm}|p{1.5cm}|}
    \hline
        \(N+1\) &\multicolumn{2}{|c|}{$S(\rho_1 \| \rho_{th})$} &\multicolumn{2}{|c|}{$T(\rho_1, \rho_{th})$}  & \multicolumn{2}{|c|}{$C_{RE}(\rho_1)$} & \multicolumn{2}{|c|}{$C_{L_1}(\rho_1)$}  \\ \hline
        3 & 4.00$\times10^{-14}$ & 2.34$\times10^{-11}$ & 7.87$\times10^{-15}$ & 2.34$\times10^{-11}$ & 5.25$\times10^{-8}$ & 2.66$\times10^{-6}$ & 8.92$\times10^{-8}$ & 5.33$\times10^{-6}$ \\ \hline
        4 & 3.03$\times10^{-11}$ & 1.15$\times10^{-8}$ & 2.95$\times10^{-11}$ & 1.15$\times10^{-8}$ & 5.81$\times10^{-5}$ & 2.05$\times10^{-6}$ & 2.98$\times10^{-6}$ & 1.16$\times10^{-4}$ \\ \hline
        5 & 1.48$\times10^{-11}$ & \(\mathbf{2.68 \times 10^{-7}}\) & 1.46$\times10^{-11}$ & \(\mathbf{2.68 \times 10^{-7}}\) & 2.05$\times10^{-6}$ & \(\mathbf{2.75 \times 10^{-4}}\) & 4.09$\times10^{-6}$ & \(\mathbf{5.51 \times 10^{-4}}\) \\ \hline
        6 & 5.59$\times10^{-15}$ & 1.09$\times10^{-8}$ & 1.64$\times10^{-15}$ & 1.09$\times10^{-8}$ & 3.03$\times10^{-8}$ & 5.48$\times10^{-5}$ & 4.02$\times10^{-8}$ & 1.10$\times10^{-4}$ \\ \hline
        7 & 7.09$\times10^{-16}$ & 6.53$\times10^{-9}$ & 1.85$\times10^{-16}$ & 6.53$\times10^{-9}$ & 5.53$\times10^{-9}$ & 4.21$\times10^{-5}$ & 1.11$\times10^{-8}$ & 8.42$\times10^{-5}$ \\ \hline
    \end{tabular}
    \label{Table::Value2dis}
\end{table*}

\section{Parameter Dependence of the Model}\label{apC}

\begin{figure*}[t!]
		\centering
		\subfloat[]{\includegraphics[width=0.45\linewidth]{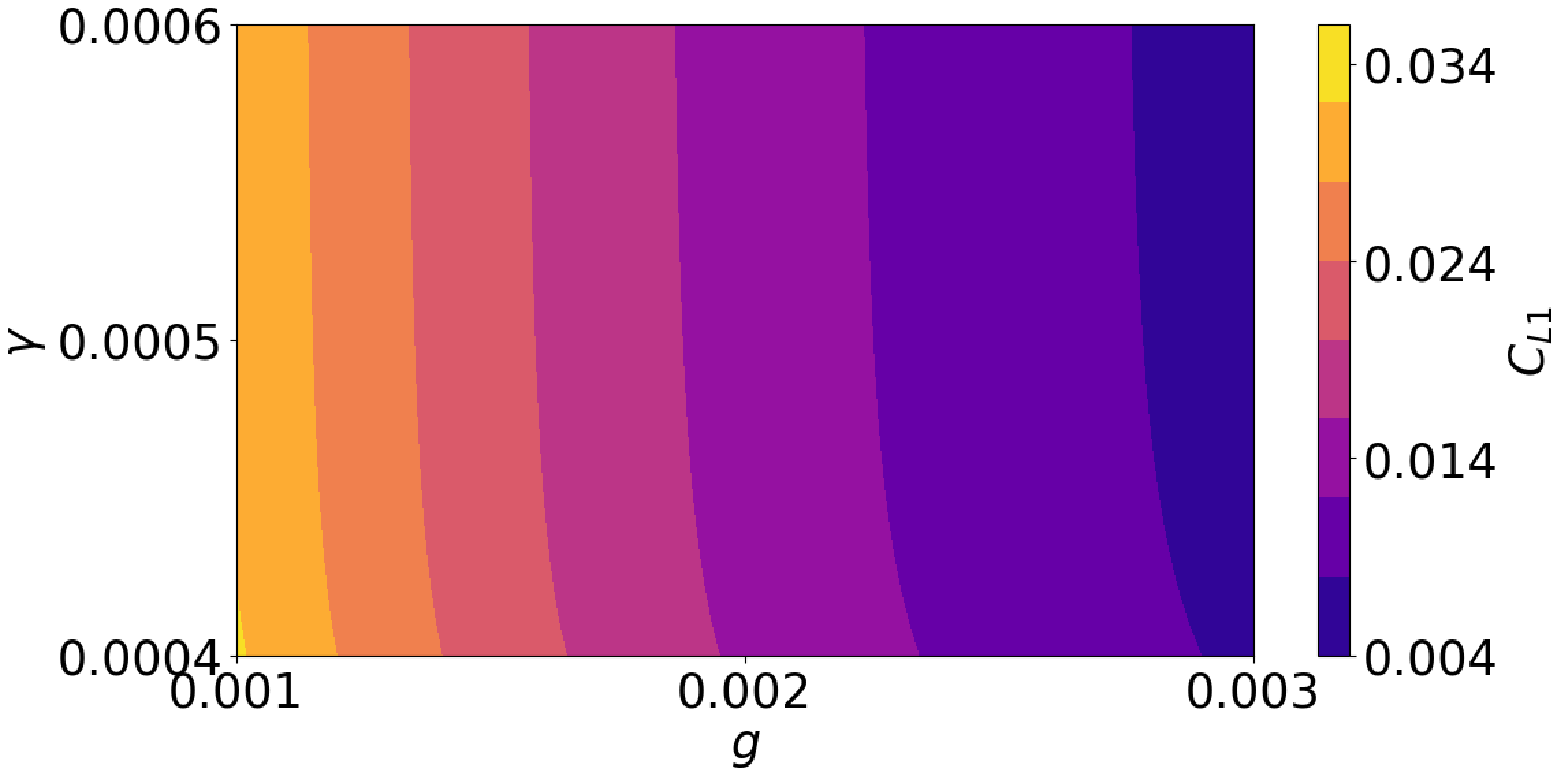}  \label{fig6:a}}
		\quad
		\subfloat[]{\includegraphics[width=0.45\linewidth]{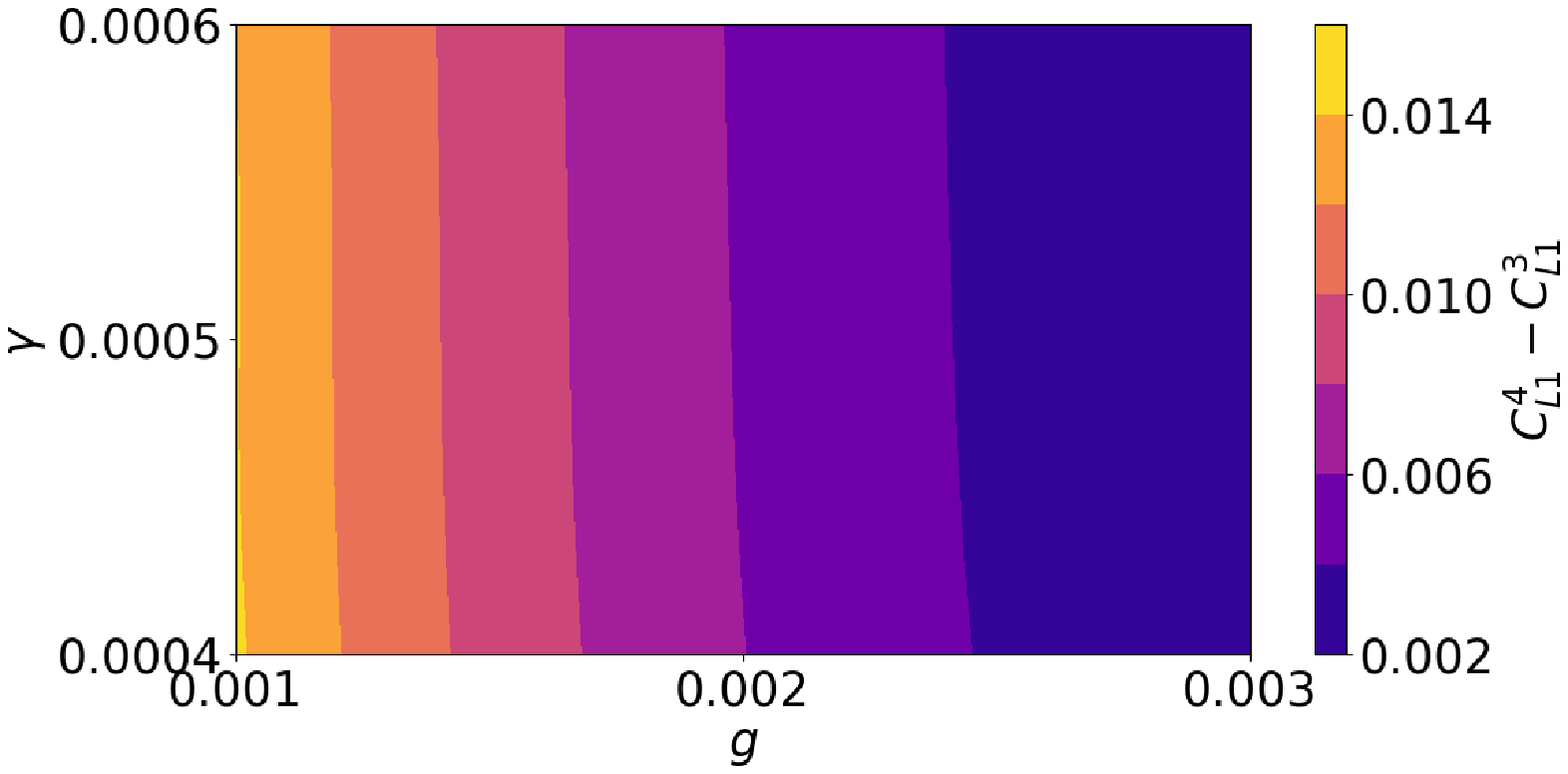}  \label{fig6:b}}
		 \caption{(a) Mean value of the L1 norm from time $t_1 = 29000 $ to time $t_2 = 30000 $ with respect to coupling constant, $g$, and dissipative constant, $\gamma$. The graph corresponds to the tetrahedral geometry (4 spin network with maximal connectivity in Fig~\ref{Fig::Geometries}). (b)  Difference of the L1 norm mean values from time $t_1 = 29000 $ to time $t_2 = 30000 $ with respect to coupling constant, $g$, and dissipative constant, $\gamma$. The graph represents the L1 norm difference between the three spin network and the four spin network with maximum connectivity (see Fig.~\ref{Fig::Geometries}). }
		 \label{Fig6}
\end{figure*}

In this paper, specific coupling constants, denoted as \(g\) for the inter-spin coupling and \(\gamma\) for coupling with the environment, were set at \(0.002\) and \(0.0005\) respectively. We further explored variations in coupling constants in the context of a four buffer spin network arranged in a tetrahedral geometry, as illustrated in Fig~\ref{fig6:a}. Interestingly, we observed that changes in these coupling constants had minimal impact on the behavior of the L1 norm. Moreover, our analysis in Fig~\ref{fig6:b} confirmed that our key findings remained consistent across different coupling scenarios. Specifically, we compared the L1 norm of the tetrahedral geometry with that of a three buffer spin network. The L1 norm of the tetrahedral geometry consistently exceeded that of the three-buffer spin network. This underscores the robustness of our main conclusion across various coupling constant settings.

\clearpage

\end{document}